\documentclass{emulateapj}

\usepackage{natbib}

\def\gtorder{\mathrel{\raise.3ex\hbox{$>$}\mkern-14mu
             \lower0.6ex\hbox{$\sim$}}}

\def\ltsima{$\; \buildrel < \over \sim \;$}
\def\simlt{\lower.5ex\hbox{\ltsima}}
\def\gtsima{$\; \buildrel > \over \sim \;$}
\def\simgt{\lower.5ex\hbox{\gtsima}}

\shorttitle{GALEX Ultraviolet Spectra of Type II-P SN 2005ay}
\shortauthors{Gal-Yam et al.}

\begin{document}


\title{GALEX Spectroscopy of SN 2005ay suggests UV spectral
uniformity among type II-P supernovae}


\author{A. Gal-Yam\altaffilmark{1,2},
F. Bufano\altaffilmark{3},
T. A. Barlow\altaffilmark{4},
E. Baron\altaffilmark{5},
S. Benetti\altaffilmark{3},
E. Cappellaro\altaffilmark{3},
P. J. Challis\altaffilmark{6},
R. S. Ellis \altaffilmark{2},
A. V. Filippenko \altaffilmark{7},
R. J. Foley\altaffilmark{7}
D. B. Fox\altaffilmark{8}, 
M. Hicken\altaffilmark{6},
R. P. Kirshner\altaffilmark{6},
D. C. Leonard\altaffilmark{9}, 
W. Li\altaffilmark{7},
D. Maoz\altaffilmark{10},
T. Matheson\altaffilmark{11},
P. A. Mazzali\altaffilmark{3,12},
M. Modjaz\altaffilmark{6,7},
K. Nomoto\altaffilmark{13},
E. O. Ofek\altaffilmark{2},
J. D. Simon\altaffilmark{2}
T. A. Small\altaffilmark{4},
G. P. Smith\altaffilmark{14},
M. Turatto\altaffilmark{3},
S. D. Van Dyk\altaffilmark{15},
and
L. Zampieri\altaffilmark{3}
}

\altaffiltext{1}{Benoziyo Center for Astrophysics, Weizmann Institute of Science, 76100 Rehovot, Israel; avishay.gal-yam@weizmann.ac.il.}
\altaffiltext{2}{Department of Astronomy, MS 105-24, California Institute of Technology, Pasadena, CA 91125.}
\altaffiltext{3}{INAF - Osservatorio Astronomico di Padova, Vicolo dell'Osservatorio 5, I-35122 Padova, Italy.}
\altaffiltext{4}{GALEX Science Center, California Institute of Technology, Pasadena, CA 91125.}
\altaffiltext{5}{Homer L. Dodge Department of Physics and Astronomy, University of Oklahoma, Norman, OK 73019-2061.}
\altaffiltext{6}{Harvard-Smithsonian Center for Astrophysics, Cambridge, MA.}
\altaffiltext{7}{Department of Astronomy, University of California, Berkeley, CA 94720-3411.}
\altaffiltext{8}{Department of Astronomy and Astrophysics, 525 Davey Laboratory, Pennsylvania State University, University Park, PA 16802.}
\altaffiltext{9}{Department of Astronomy, San Diego State University, San Diego, CA.}
\altaffiltext{10}{School of Physics and Astronomy, Tel Aviv University, 69978 Tel Aviv, Israel}
\altaffiltext{11}{National Optical Astronomy Observatory, Tucson, AZ 85719.}
\altaffiltext{12}{Max-Planck-Institut f\"{u}r Astrophysik, Karl-Schwarzschild-Str. 1, 85748 Garching, Germany}
\altaffiltext{13}{Institute for the Physics and Mathematics of the Universe, and
Department of Astronomy, University of Tokyo, Chiba 277-8582, Japan}
\altaffiltext{14}{School of Physics and Astronomy, University of Birmingham, Edgbaston, Birmingham B15 2TT, UK}
\altaffiltext{15}{Spitzer Science Center, California Institute of Technology, Mail Code 220-6, 1200 East California Boulevard, Pasadena, CA 91125}



\begin{abstract}

We present the first results from our GALEX program designed to obtain
ultraviolet (UV) spectroscopy of nearby core-collapse supernovae (SNe). 
Our first target, SN 2005ay in the nearby galaxy NGC 3938, is a typical
member of the II-P SN subclass. Our spectra show remarkable similarity
to those of the prototypical type II-P event SN 1999em, and resemble
also {\it Swift} observations of the recent type II-P 
event SN 2005cs. Taken together, the observations of these three events
trace the UV spectral evolution of SNe II-P during the first month after
explosion, as required in order to interpret optical observations 
of high-redshift SNe II-P, and to derive cross-filter K-corrections.
While still highly preliminary, the apparent UV homogeneity of SNe II-P
bodes well for the use of these events as cosmological probes at high
redshift. \\

\end{abstract}


\keywords{supernovae: individual (SN 2005ay) -- ultraviolet: general}


\section{Introduction}

In order to interpret optical observations of high-redshift supernovae (SNe), 
sampling the restframe ultraviolet (UV) radiation of these events, we need to 
have UV observations of local SNe of all types. Studies of high-redshift SNe promise, in turn,
to shed light on key open questions, such as the evolution of cosmic 
metallicity, star formation at high redshift, and SN ``feedback'' processes
shaping galaxy formation. 

Type Ia SNe, famed for
their cosmological utility as precision distance estimators, are the
best-studied of all SN subclasses in restframe UV (e.g., Kirshner et al. 1993;
Ellis et al. 2008; Foley et al. 2007, 2008). SNe Ia are broadly thought to result from 
thermonuclear explosions of white dwarf stars approaching the critical
Chandrasekhar mass due to accretion from (or a merger with) a binary
companion, and show remarkable homogeneity in their observational
properties. However, UV studies may hint at unexplained dispersion in the
restframe UV band (Ellis et al. 2008). 

All other types of SNe (see Filippenko 1997 for a review) 
probably result from the gravitational collapse of 
short-lived massive stars (e.g., Crockett et al. 2008; Li et al. 2007, 
Gal-Yam et al. 2007 and references therein). In general, core-collapse SNe
are extremely heterogeneous in every observational respect, and, in 
particular, different types of core-collapse events have diverse UV properties
(e.g., UV-bright type IIn SN 1998S, Lentz et al. 2001 vs. UV-deficient type 
Ic SN 1994I, Millard et al. 1999). The dispersion in UV properties among objects within
specific core-collapse SN subtypes are so far unknown. 

Unfortunately, UV spectroscopy of reasonable quality was obtained
for only a handful of core-collapse SNe (Panagia 2003; 2007 for reviews)
and some of the best-observed events (notably SN 1987A, e.g.,
Eastman \& Kirshner 1989) are quite peculiar. This
observational deficit introduces significant uncertainties into the interpretation
of high-redshift SN observations, which are forced to rely either
on little-tested models for the UV spectrum of core-collapse events, or 
on the use of the few observed UV spectra for the entire population,
neglecting possible dispersion in spectral evolution. 
The sparse database of UV core-collapse SN spectroscopy 
continues to limit the scientific utility of large samples 
of core-collapse SNe at high redshifts.
These include both samples currently assembled using the 
{\it Hubble Space Telescope} ({\it HST}; e.g., GOODS, Dahlen et al. 2004; 
Riess et al. 2007) and from the ground (e.g., Poznanski et al. 2007a),
and future datasets expected to emerge from a possible JDEM mission that
includes a SN component and from deep infrared observations
with the {\it James Webb Space Telescope}. Furthermore,
in order to probe the physics of core-collapse SNe, broad spectral coverage, 
particularly of the UV range, where line blanketing by iron peak elements 
plays a crucial role in the formation of the spectrum (e.g., Pauldrach et al. 1996), 
is essential. UV spectra are most sensitive to the metal content of the SN ejecta, a key probe 
of the nucleosynthetic evolution.

To alleviate this problem is
the main motivation for our target of opportunity GALEX program (GALEX-GI-44, cycle 1;
GALEX-GI-67, cycle 2; GALEX-GI-61, cycle 3; GALEX-GI-20, cycle 4; PI Gal-Yam), designed to obtain 
multi-epoch UV spectroscopy of nearby core-collapse SNe. Here, we report
the first results from this program. In $\S~2$ we describe our observation,
and in $\S~3$ we present our results. Discussion and conclusions follow
in $\S~4$. UT dates are used throughout the paper.

\section{Observations}

SN 2005ay in the nearby galaxy NGC 3938 (heliocentric velocity $v=809~ {\rm  km~s}^{-1}$)
was discovered on 2005 March 27 by Rich (2005) and rapidly identified as a type II supernova
(Taubenberger et al. 2005; Gal-Yam \& Smith 2005). Pre-explosion limits (2005 March 19; Yamaoka \& Itagaki 2005)
indicate it was discovered shortly after explosion, leading us to trigger our GALEX target 
of opportunity program.  

Following the activation of our GALEX program, we launched an intensive campaign to study
SN 2005ay from the ground in optical and IR wavelengths, to complement the GALEX UV study. 
We obtained dense multicolor photometry during the months after the SN discovery, and followed
its decline for approximately a year (Fig. 1). Numerous optical spectra follow
the evolution of this object in great detail, starting shortly after explosion (Fig. 2). 
Our observations confirm the identification of this event as a typical SN II-P (Tsvetkov et al. 2006),
and will be reported in detail by Bufano et al. (in preparation). In this {\it Letter} we 
present the earliest GALEX UV spectra of SN 2005ay and compare it with additional examples from the 
literature (Fig. 3). The spectral similarity between these supernovae
highlights an emerging picture of uniform UV properties of SNe II-P,
the most common subclass of core-collapse SNe, and the first for which analysis based on a
sample of events is possible. 

The GALEX mission observed SN 2005ay in grism spectroscopy mode on 2005 April 2, 3, 8, 15 and 16. 
At that time, only the near-UV (NUV) camera was operational, yielding a wavelength coverage of
$1800-2900$~\AA. Reduction of these data required a custom approach, in order to properly remove the
contribution from UV-bright knots and residual small-scale structure in the nearby spiral arm of the
host galaxy, NGC 3938. This process will be presented in detail by Bufano et al. (in preparation). 
In brief, we have obtained reference grism-dispersed images of the location
of SN 2005ay on 2007 March 4-28, presumably including only negligible light from
SN 2005ay, using an identical instrumental configuration. We performed digital image subtraction using the
2007 epoch as reference, and extracted our spectra from the background-subtracted reference frames. 
Fig. 3 shows the combined spectra obtained on Apr. 2, 2005 (4 GALEX orbits; total exposure
time 5705 s) and Apr. 3 2005 (7 orbits; 9516 s). The spectra were binned to $30$~\AA~
resolution elements to increase the signal-to-noise ratio. 
Errors for each spectral bin were directly measured from
the dispersion among consecutive orbits (1$\sigma$ errors are plotted in Fig. 3). 
  
\begin{figure}
\label{fig_lc}
\includegraphics[width=8.5cm]{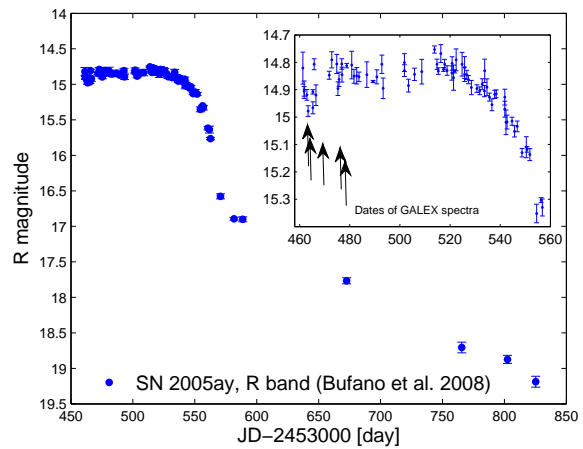}
\caption{$R$-band light curve of SN 2005ay, showing the classical features
of a type II-P event, including the initial extended plateau, followed by a rapid
decline period, and terminating in a radioactive decline phase. The GALEX data were
obtained in the first two weeks of the plateau (see inset). The full multiband 
light curve data, along with a detailed discussion of their calibration and 
error analysis, will be presented by Bufano et al. (in preparation). 
}   
\end{figure}

\begin{figure}
\includegraphics[width=8.5cm]{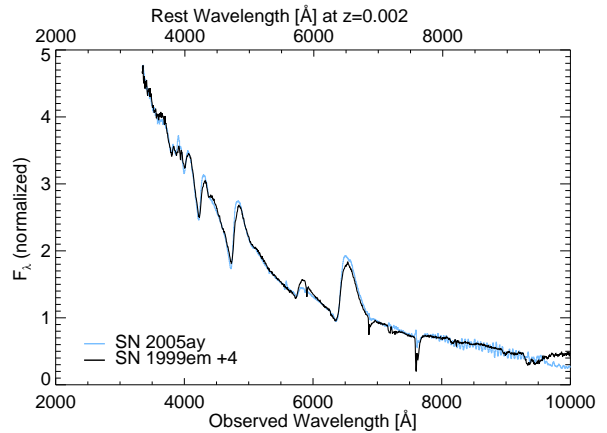}
\caption{An optical spectrum of SN 2005ay obtained on Mar 31 
using the Dolores spectrograph mounted on the 3.5m TNG telescope
in La Palma, Spain. Comparison with
a database of SN spectra using the {\it Superfit} code (Howell et al. 2005) 
reveals a remarkable similarity to early spectra of the prototypical
SN II-P event, SN 1999em (Leonard et al. 2002). The fit presented 
requires the application of a small negative reddening correction ($A_V\sim0.05$, indicating 
that SN 2005ay probably had lower dust extinction than SN 1999em, which itself
was only slightly extinguished ($A_V\approx0.35$ mag; Leonard et al. 2002). 
Alternatively, this could indicate a slightly higher temperature 
for SNe 2005ay. The blue 
continuum and similarity to very early spectra of SN 1999em confirm the imaging 
indications that this object was discovered shortly (a few days at most) after
explosion. Additional data and further discussion will be presented by Bufano
et al. (in preparation). 
}   
\label{fig_spec}
\end{figure} 

\begin{figure}
\begin{center}
\label{fig_uvspec}
\includegraphics[width=8.5cm]{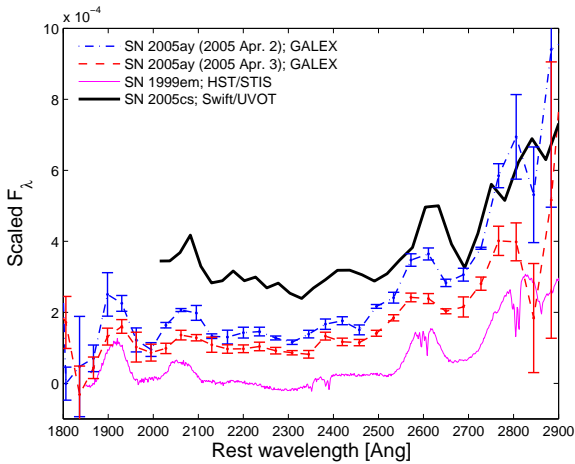}
\caption{
Comparison of the restframe-UV spectra of type II-P 
SN 1999em ({\it HST}/STIS, magenta; Baron et al. 2000), SN 2005cs ({\it Swift}/UVOT,
black; Brown et al. 2007), and our new GALEX spectra of SN 2005ay (red, blue). 
The spectra of SNe 1999em and 2005ay were obtained approximately 12 days after
explosion. The {\it Swift} spectrum of SN 2005cs was obtained somewhat earlier (6 days
post-explosion) and we have adjusted its shape to match the spectral evolution 
of the SNe to day 12, 
so it should be considered approximate only; see text for details.  
Note the remarkable similarity in both continuum shape and observed features
among these three events. 
}   
\end{center}
\end{figure}

\section{Discussion}

\subsection{UV Homogeneity}

In Figure 3, we compare our restframe-UV spectra of SN 2005ay with
those of two other SNe II-P, SN 1999em and SN 2005cs, which are the only other 
II-P events with high signal-to-noise ratio restframe-UV spectra. SN 1999em
was discovered on 1999 October 28.9, and the estimated explosion
date is $\sim5.3$ days before discovery (Leonard et al. 2000). The {\it
HST} spectrum reproduced here (Baron et al. 2000) 
was obtained on 1999 Nov. 5, i.e., approximately
12 days after explosion. SN 2005cs was discovered on 2005 June 29, approximately
two days post-explosion (Dessart et al. 2008). The {\it Swift} spectrum reproduced here
was obtained on 2005 July 3, i.e., approximately 6 days post-explosion (Brown
et al. 2007). Tsvetkov et al. (2006) estimate the explosion date of 
SN 2005ay as 2005 March $23\pm2$ days. From spectral comparison with 
SN 1999em (Fig. 2) we find that the TNG spectrum of SN 2005ay has, 
consistently, an age of approximately 9 days after explosion. So our GALEX
spectra reported in Fig. 3 were obtained 11-13 days post explosion. 
This estimate is also consistent with available pre-explosion limits ($\S~2$). 

Since the UV spectra of SNe evolve rapidly, in order to compare our GALEX spectra
with those from the literature we had to adjust the {\it Swift} spectrum of SN 2005cs 
(obtained $\sim6$ days post explosion) to approximately match those of the other events,
both of which happen to have been obtained at a similar epoch (11-13 days post explosion). 
We perform this adjustment using two alternative methods, giving consistent results.
First we interpolated the {\it {\it Swift}} photometry reported by Brown et al. (2007)
and Dessart et al. (2008) in the $U$, $UVW1$ and $UVW2$ bands (central wavelengths
$3465$ \AA, $2600$ \AA, and $1928$ \AA, respectively, see Brown et al. 2007 for 
additional bandpass details), which bracket our wavelength range, to 6 and 12 days
post explosion. We then adjust the spectral shape of the early spectrum to match 
the evolution of the spectral energy distribution of the SN, measured from
{\it Swift} photometry. 

Alternatively, we can use an additional spectrum obtained by {\it Swift} $\sim11$ days after
explosion. This later spectrum has much lower signal-to-noise ratio (see Brown et al. 2007).
However, assuming that the spectral shape of this late spectrum is correct, even if 
individual features cannot be reliably measured, we can adjust the day 6, high signal-to-noise
spectrum to match the shape of the day 11 spectrum. This yields consistent
results with those obtained using the UV photometry (see above) and we plot the adjusted 
spectrum in black in Figure 3. A more detailed discussion of the {\it Swift} spectra
of SN 2005cs will appear in Bufano et al. (in preparation).   
Obviously, both these procedures
introduce additional uncertainties to the spectral shape of SN 2005cs, which
should therefore be regarded as approximate only (in contrast to the other spectra in Fig. 3 
that have not been modified from their observed form). 

UV spectra are particularly sensitive to effects of dust extinction. Since the estimated
extinction for the three events in question is uncertain but small ($E_{B-V}\lesssim0.1$ mag; Leonard et al.
2002, Baron et al. 2000; Pastorello et al. 2006; Brown et al. 2007; Dessart et al. 2008; 
Tsvetkov et al. 2006; this work (Fig. 2)), we have made no correction for extinction to the spectra
in Fig. 3, and postpone a more detailed discussion of this issue to Bufano et al. (in preparation).  

Inspecting Fig. 3, we note the remarkable similarity among the spectra of these three SNe.
Identical spectral features are prominent, i.e., the Mg II P-Cygni profile around
$2800$ \AA, as well as the emission ``bumps'' around $2600$ \AA, $2200$ \AA, $2070$ \AA,
and $1900$ \AA. This similarity is independent of the overall spectral shape and suggests that
similar physical conditions occur at the photospheres of these SNe. 
The continuum shape is also almost identical when SN 1999em and
SN 2005ay are compared, with SN 2005cs also a close match, considering the additional
uncertainty introduced to compensate for the earlier observation (see above). This
apparent homogeneity is in stark contrast to the wide diversity among UV spectra 
of other subtypes of core-collapse SNe, and is instead reminiscent of the uniformity
among UV spectra of standard type Ia supernovae (e.g., Foley et al. 2008). 

The combined set of restframe UV spectroscopy now available for SNe II-P 
(SN 1999em, SN 2005cs and SN 2005ay) samples the UV spectral evolution of
such events between 6 days (the earliest {\it Swift} observation of SN 2005cs)
and 25 days (our last GALEX spectrum of SN 2005ay) after explosion. Assuming these
events are representative of the population of SNe II-P, the data suffice to
describe the UV evolution these objects throughout the interesting UV-bright period
after explosion (Bufano et al., in preparation).  

\subsection{Implications}

The use of SNe II-P as an alternative cosmological distance indicator to SNe Ia
has been advocated for many years (e.g., Kirshner \& Kwan 1974; Wagoner 1977; Schmidt et al. 1994;
Hamuy \& Pinto 2002; Baron et al. 2004). Nugent et al. (2006) have recently
demonstrated that SN II-P distances can be measured, with existing facilities,
out to cosmological distances, and that an independent cosmological measurement
is feasible. A possible advantage of SNe II-P for cosmological use is that these
events, arising from the explosion of relatively low-mass red supergiant stars
(e.g., Hendry et al. 2006; Li et al. 2007 and references therein) 
are much more common (per unit volume) than SNe Ia, and should occur
in abundance out to high redshifts, where such massive stars are 
formed in great numbers. SNe Ia, in contrast, may show a decline in rate
above $z\approx1.5$ (Dahlen et al. 2008; though see Poznanski et al. 2007a), which,
if real, may reflect a metallicity ``floor'' required for these events
to take place (Kobayashi et al. 1998), or a long delay time between the formation
of SN Ia progenitors and their explosion (Strolger et al. 2004). 

Since optical detectors still offer the most powerful combination of efficiency, low
sky background, and 
relatively wide fields, taking advantage of the benefits of SNe II-P for cosmology
at high redshifts may require the use of optical surveys, sampling the restframe
UV, to discover high redshift SNe II-P. A UV spectral homogeneity as suggested by our data,
if confirmed, would allow the use of UV spectral templates to discover and photometrically
identify high-redshift SNe II-P (e.g., Poznanski et al. 2002; Riess et al. 2004; 
Poznanski et al. 2007b). Furthermore, accurate photometry could be derived using 
cross-filter K-corrections which take into account the effects of 
spectral features using template spectra (S-corrections; Stritzinger et al. 2002).

Another important application of high-redshift supernova surveys is to measure supernova
rates, which probe the star formation rate and cosmic metal production. 
SNe that decline quickly in the UV would be 
detectable for shorter periods in restframe-UV surveys, and thus a small number
of detections is translated into a higher true rate, while a smaller
correction needs to be applied to SNe that remain bright in restframe UV longer.
Properly calculating these corrections for SNe at different redshifts, for which
the same survey band samples different spectral ranges within the restframe UV,
requires detailed knowledge about the spectral evolution of SNe, one of 
the goals of our GALEX program.  
As shown here, for SNe II-P, the most common type of core-collapse
subtypes, we have made good progress in achieving this goal. 

\section{Conclusions}

We are carrying out a spectroscopic survey of nearby core-collapse SNe using 
GALEX grism spectroscopy in target-of-opportunity mode. About $1-2$ nearby events
are observed each year, increasing our knowledge of the spectral evolution
of core-collapse SNe of the various subtypes. Our collaboration also provides
supporting IR and optical observations of our GALEX targets. 
We have presented first results from this project -- spectra of the nearby 
type II-P SN 2005ay. Combined with additional observations of two similar 
objects from the literature, we trace the UV spectral evolution of SNe II-P
and find a remarkable similarity among these objects, 
the most common type of core-collapse SNe and the only subtype
with a sample of events having UV spectroscopic measurements. 
Such restframe-UV homogeneity, if supported by additional objects, indicates that
the use of these SNe as cosmological probes is a promising prospect.  

\section*{Acknowledgments}
Based on observations made with the NASA Galaxy Evolution Explorer,
GALEX, which is operated for NASA by the California Institute of Technology
under NASA contract NAS5-98034. We further acknowledge financial support
from NASA through the GALEX guest investigator program (projects 
GALEX-GI-44, cycle 1; GALEX-GI-67, cycle 3, and GALEX-GI-20, cycle 4).
We are indebted to the GALEX Science Operations Center (SOC), and in particular to K. Forster, for
making this ToO program possible. We thank D. Poznanski for comments
on the manuscript.   
A.G. acknowledges support by the Benoziyo Center for Astrophysics
and the William Z. and Eda Bess Novick New Scientists Fund at the
Weizmann Institute. 
SB, EC and MT are supported by the Italian Ministry of Education
via the PRIN 2006 n.2006022731 002.

\clearpage


\end{document}